\documentclass[aip,preprint]{revtex4-2}
\usepackage{tikz}
\usepackage{pgfplots}
\usepackage{pgfplotstable}
\pgfplotsset{width=10cm,compat=1.9}
\usepackage[cp1251]{inputenc}
\usepackage[T2A]{fontenc}
\usepackage[english]{babel}
\usepackage{amssymb,latexsym,amsmath,amscd}
\usepackage{graphicx,color,framed}
\usepackage{bm}
\usepackage{appendix}
\usepackage{tabu}
\usepackage{tikz}
\usepackage{xcolor}
\usepackage{siunitx}
\usepackage{empheq}
\usepackage{float}
\usepackage{subcaption}
\usepackage{xr}
\usepackage{siunitx}
\usepackage{microtype}
\usepackage{amsfonts}
\usepackage{gensymb}
\usepackage[normalem]{ulem}
\graphicspath{{figures/}}

\makeatletter
\newcommand*{\addFileDependency}[1]{% argument=file name and extension
  \typeout{(#1)}
  \@addtofilelist{#1}
  \IfFileExists{#1}{}{\typeout{No file #1.}}
}
\makeatother

\begin{document}

%\homepage[]{Your web page}
%\thanks{}
%\altaffiliation{}
\author{\firstname{Nikolai N.} \surname{Kalikin}}
\affiliation{Laboratory of Multiscale Modeling of Molecular Systems, G.A. Krestov Institute of Solution Chemistry of the Russian Academy of Sciences, 153045, Akademicheskaya st. 1, Ivanovo, Russia}
\author{\firstname{Yury A.} \surname{Budkov}}
\email[]{ybudkov@hse.ru}
\affiliation{Laboratory of Multiscale Modeling of Molecular Systems, G.A. Krestov Institute of Solution Chemistry of the Russian Academy of Sciences, 153045, Akademicheskaya st. 1, Ivanovo, Russia}
\affiliation{Laboratory of Computational Physics, HSE University, Tallinskaya st. 34, 123458 Moscow, Russia}
\affiliation{School of Applied Mathematics, HSE University, Tallinskaya st. 34, 123458 Moscow, Russia}
%\homepage[]{Your web page}
%\thanks{}
%\altaffiliation{}
\title{Modified Debye-H\"{u}ckel-Onsager Theory for Electrical Conductivity in Aqueous Electrolyte Solutions. Account of Ionic Charge Nonlocality}
\begin{abstract}
The paper presents a mean field theory of electrolyte solutions, extending the classical Debye-H\"{u}ckel-Onsager theory to provide a detailed description of the electrical conductivity in strong electrolyte solutions. The theory systematically incorporates the effects of ion specificity, such as steric interactions, hydration of ions, and their spatial charge distributions, into the mean-field framework. This allows for calculation of ion mobility and electrical conductivity, while accounting for relaxation and hydrodynamic phenomena. At low concentrations, the model reproduces the well-known Kohlrausch's limiting law. Using the exponential (Slater-type) charge distribution function for solvated ions, we demonstrate that experimental data on the electrical conductivity of aqueous 1:1, 2:1, and 3:1 electrolyte solutions can be approximated over a broad concentration range by adjusting a single free parameter representing the spatial scale of the nonlocal ion charge distribution. Using the fitted value of this parameter at 298.15 K, we obtain good agreement with available experimental data when calculating the electrical conductivity across different temperatures. We also analyze the effects of temperature and electrolyte concentration on the relaxation and electrophoretic contributions to total electrical conductivity, explaining the underlying physical mechanisms responsible for the observed behavior.
\end{abstract}

\maketitle

\section{Introduction}
Electrolyte solutions~\cite{robinson2002electrolyte} are essential for maintaining proper function and balance within living organisms, as they contain ions crucial for cellular processes such as nerve signaling~\cite{cooke1994neuroimmune,schoen2007mechanism}, muscle contraction~\cite{bohr1958effect,bohr1964electrolytes}, and fluid balance~\cite{metheny2012fluid}. These solutions can be found in natural sources like fruits and vegetables, as well as in sports drinks and oral rehydration solutions. Beyond biology and the food industry, electrolytes are critical components in technologies with significant societal and environmental impacts. In battery technology, electrolytes facilitate the movement of ions between electrodes, allowing energy storage and release~\cite{cussen2010structure,zahn2016improving,deng2017enhancing,long2016polymer,szczkesna2023lithium}. With growing global demand for fresh water, the role of electrolytes in desalination processes~\cite{alkhadra2022electrochemical,chaplin2019prospect} is becoming increasingly important. Electrolytes play a key role in technologies such as reverse osmosis and electrodialysis, facilitating the removal of contaminants from water and the production of clean drinking water. Thus, predicting electrical conductivity in solutions containing strong electrolytes poses significant challenges in physical chemistry and chemical physics, with implications for modern biology, medicine, food industry, agriculture, electronics, water purification, and environmental monitoring technologies.

Debye, H\"{u}ckel, Onsager, Fuoss, and Falkenhagen~\cite{debye1987,debye1923theory,debye1928dispersion,onsager1927report,onsager1927theory,onsager2002irreversible,fuoss1978review} laid the foundation for understanding the {microscopic} theory of electrical conductivity in electrolytes. More recently, renewed interest and progress in this field have arisen from advancements in the statistical theory of liquids and liquid-phase solutions since the publication of these seminal papers. Approaches based on integral equations for distribution functions~\cite{bernard2023analytical,naseri2023new,boroujeni2023estimation}, stochastic density functional theory (SDFT)~\cite{demery2016conductivity,avni2022conductivity_,avni2022conductance,bernard2023analytical}, and self-consistent field theory based on the Poisson-Boltzmann equation~\cite{vinogradova2023electrophoresis,vinogradova2023conductivity} have been developed and applied with varying degrees of success to model the electrical conductivity of electrolyte solutions across a broad range of concentrations and temperatures.

Let us review the current state of the art in this field. {In paper~\cite{fraenkel2018improved}, the author proposed a modified Debye-H\"uckel-Onsager (DHO) theory, using the concept of smaller-ion shells, which was initially developed by the same author in~\cite{fraenkel2010simplified}, as a way to enhance the accuracy of predicting activity coefficients for ions. The author's estimation of the electrical conductivity and activity coefficients of numerous aqueous electrolyte solutions with varying valences, up to concentrations of 0.5-1 mol/l, was achieved without the use of any adjustable parameters and is considered accurate.} The paper~\cite{avni2022conductivity_} examines the conductivity of ionic solutions using SDFT combined with a modified Coulomb interaction that takes into account the hard-core repulsion between ions. The authors established that the modified potential suppresses short-range electrostatic interactions present in the original Debye-H\"uckel-Onsager theory, leading to improved conductivity results. The authors also claim that the conductivity values obtained from the model agree well with the experimental data up to a concentration of 3 mol/l. In a recent paper~\cite{avni2022conductance}, the same authors extended their study to include multivalent ions and high electric fields, where a deviation from Ohm's law, known as the Wien effect, occurs. In article~\cite{bernard2023analytical}, the authors explore recent approximations for computing the conductivity and self-diffusion coefficient of electrolytes, with a focus on the specifics of truncating Coulomb interactions at short distances. They compare these new approaches to conventional methods, such as the mean-spherical approximation and mode-coupling theory. The authors also discuss the importance of considering hydrodynamic effects in SDFT and demonstrate how the choice of modified Coulomb interactions can affect the properties of electrolytes. In general, the authors highlight the need for continued development and refinement of SDFT approaches for studying electrolytes and emphasize the valuable insights that can be gained by comparing these methods with other theoretical frameworks. In study~\cite{naseri2023new}, the authors developed a new electrical conductivity model for non-associated electrolyte solutions based on the Debye-H\"{u}ckel-Onsager theory. This model assumes a single cation and a single anion within a continuum medium of the solvent(s), utilizing their crystallographic ionic radii. The authors compared the predictions of their model with experimental measurements of various binary aqueous solutions over a wide temperature range, from 273.15 to 373.15 K, and demonstrated good agreement with the experimental data. In paper~\cite{vinogradova2023electrophoresis}, the authors provided a comprehensive examination of the unique behavior of electrolyte solutions conductivity when confined to micro- and nanochannels, as compared to bulk phases. They focuses on the reduction of ion mobility with increasing salt concentration, which impacts the conductivity of monovalent solutions as they transition from bulk to confinement in a narrow slit. The authors systematically treated electrophoresis of ions, deriving equations for zeta potentials and ion mobilities, and subsequently derived an elegant analytical expression for bulk conductivity. In the paper \cite{vinogradova2023conductivity}, the same authors demonstrated that the derived expression accurately describes the conductivity of univalent salt solutions. Note that this model takes into account only a mean-field approximation of electrostatic interactions and uses an adapted classical approach to calculate colloid electrophoresis, as described in~\cite{henry1931cataphoresis} and~\cite{maduar2015electrohydrodynamics}. It is interesting to note that the authors reached an excellent agreement with the experiment without considering the relaxation effect of the salt ions.

Despite several attempts to theoretically describe electrical conductivity in aqueous solutions of strong electrolytes, some important issues remain unclear from the perspective of chemical physics and physical chemistry. \textbf{Firstly}, to the best of our knowledge, the nonlocal nature of the ion charge distribution has not been systematically explored in relation to the electrical conductivity of electrolyte solutions. More specifically, according to quantum theory, the charge of solvated ions (with surrounding solvent molecules in solvation shell) is distributed over a region rather than being concentrated at a single point or spread over a thin shell. In other words, from the perspective of electrostatics, it should behave more like a smooth cloud than a point-like charge or a uniformly charged spherical shell. Based on general considerations, this effect is expected to become significant at sufficiently high solution concentrations, where the interactions between ions can no longer be accurately described as interactions between the point-like charges. It should be noted that recent studies have effectively attempted to account for charge nonlocality by modifying the Coulomb potential at short distances~\cite{avni2022conductance,avni2022conductivity_,bernard2023analytical}. However, these modifications have not yet been explicitly related to the nature of the charge distribution nonlocality. Note that renormalization of the Coulomb potential at short distances has been systematically considered in plasma physics taking into account quantum effects related to the Heisenberg uncertainty relation~\cite{kraeft1986quantum}. Quantum mechanical calculations of the effective electrostatic interactions between ions in electrolyte solutions present a greater challenge than those in plasma physics, primary due to the presence of multiple electrons in solvated ions. To address this task one can resort to computationally expensive {\sl ab initio} molecular dynamics simulations. However, for practical calculations, it is also possible to use phenomenological models of the charge distribution of solvated ions adopted in different phenomenological theories~\cite{dogonadze1974polar,kornyshev1996shape,rubashkin2015nonlocal,de2020interfacial} or within the microscopic statistical field theory~\cite{lue2006variational,wang2010fluctuation,budkov2020statistical,budkov2023variational,nikiforova2024modeling}. These approaches lead to a truncation of the Coulomb potential at short distances. \textbf{Secondly}, the effect of ion hydration on the phenomenon of ion transport in a solution under an external electric field is not fully understood. This effect should be analyzed by considering the effective size of the ions (ion hydration radii), taking into account the steric interactions between the ions and their effect on the dielectric constant of the solution (dielectric decrements~\cite{nakayama2015differential,ben2011dielectric,mazurunderstanding}). For this purpose, steric interactions between ions at close distances could be modeled as interactions of hard spheres with diameter equal to twice the hydration (solvation) radius. It is important to note that existing theoretical models~\cite{bernard2023analytical,avni2022conductance,avni2022conductivity_} take into account steric effects using crystallographic radii of ions, rather than solvation radii.

This study proposes a modified version of the Debye-H\"{u}ckel-Onsager (DHO) theory that takes into account ion solvation, steric interactions between ions, and the non-local distribution of charge on solvated ions. We utilize this theory to derive analytical expressions for the electrophoretic and relaxation contributions to ion mobility, as well as to describe the electrical conductivity of electrolyte solutions containing ions of different valencies across a wide range of concentrations and temperatures.

\section{Statement of problem}
It is well known~\cite{pitaevskii2012physical} that in solutions of strong electrolytes in an electric field $\bold{E}$, each ion experiences a force of $q_{a}\bold{E}$, resulting in an average directed velocity of $\bold{v}_a=m_aq_{a }\bold{E}$ for the ion, where $m_a$ represents the mobility of the ion of type $a$. The current density in the solution can be expressed as
\begin{equation}
\bold{j}=\bold{E}\sum\limits_{a}q_{a}^2c_{a}m_a,
\end{equation}
where $c_{a}$ denotes the concentration of ions of type $a$ in the bulk solution. Therefore, the conductivity is given by the relation
\begin{equation}
\sigma =\sum\limits_{a}q_{a}^2c_{a}m_a.
\end{equation}

For a very diluted solution, the mobility of individual ions approaches a constant limit given by the equation:
\begin{equation}
m_a^{(0)}=\frac{1}{\gamma_a}.
\end{equation}
For spherical ions, the viscous friction coefficient can be expressed as $\gamma_{a}=6\pi\eta R_a$, where $R_a$ represents the hydrodynamic (Stokes) radius of the $a$-th ion, and $\eta$ denotes the shear viscosity of the solution. In solutions at finite concentrations each ion is surrounded by a screening charge cloud (atmosphere), which alters its mobility due to two distinct effects. The motion of ions in an external field causes a distortion in the charge distribution within the cloud, leading to an additional 'relaxation' field acting on the ion. Furthermore, the movement of the ionic atmosphere induces motion in the solvent, resulting in the ion being "carried off". The impact of the first type of contribution on mobility is described as {\sl relaxation}\cite{pitaevskii2012physical,pitts1953extension,fuoss1978review}, while the second type is referred to as {\sl electrophoretic}~\cite{pitaevskii2012physical,pitts1953extension,vinogradova2023electrophoresis}. In the following, we will present an approach that extends the original Debye-H\"{u}ckel-Onsager theory~\cite{fuoss1978review} to calculate both contributions to the electrical conductivity of strong electrolyte solutions.

\section{Relaxation contribution}
Let us begin by calculating the relaxation contribution to the ion mobilities. We can define the two-point ion concentrations, denoted as $c_{ab}(\bold{r}_a,\bold{r}_b)$. The value $c_{ab}(\bold{r}_a,\bold{r}_b)\text{d}\bold{r}_a$ represents the number of ions of type $a$ near point $\bold{r}_a$ when a test ion of type $b$ is fixed at point $\bold{r}_b$. It is important to note that the ions can be of the same or different types, so we have the following symmetry relation:
\begin{equation}
\label{trans}
c_{ab}(\bold{r}_a,\bold{r}_b)=c_{ba}(\bold{r}_b,\bold{r}_a).
\end{equation}
As the distance $|\bold{r}_a-\bold{r}_b|\to \infty$, the function $c_{ab}(\bold{r}_a,\bold{r}_b)$ approaches a bulk value $c_{a}$. Thus, we can relate the two-point concentration $c_{ab}$ with the standard pair correlation function, $g_{ab}$, i.e. $c_{ab}(\bold{r}_a,\bold{r}_b)=c_{a}g_{ab}(\bold{r}_a,\bold{r}_b)$.  In the absence of an external field $\bold{E}$, the function $c_{ab}$ becomes dependent only on the distances $|\bold{r}_a-\bold{r}_b|$. This dependency is not present when the external field is applied ($\bold{E} \neq 0$). We take into account the spatial charge distribution of solvated ions. This can be done by introducing charge distribution functions $\varrho_a(\bold{r})$, which, in the general case, can be obtained from quantum mechanical calculations. However, for practical estimates, they can be treated as phenomenological functions. In what follows, we assume for all ionic species $\varrho_a(\bold{r})=\varrho(\bold{r})$. This simplification will allow us to obtain more compact analytical expressions. As will be demonstrated below, the electrical conductivity of most one-component aqueous electrolyte solutions can be described using a model with equal charge form factors.

\textbf{Modified Onsager-Fuoss equation.} In the configuration space of two ions, $a$ and $b$, in a steady state, which we will only consider in this work, the following continuity equation holds~\cite{pitaevskii2012physical,onsager2002irreversible,fuoss1978review}
\begin{equation}
\label{continous}
\nabla_a\cdot \bold{j}_a +\nabla_b\cdot \bold{j}_b =0,
\end{equation}
where $\bold{j}_{a}$, $\bold{j}_{b}$ are the particle flux densities for ions of types $a$ and $b$, respectively. The subscripts $a$ and $b$ on the operator $\nabla$ indicate a differentiation with respect to the coordinates of the corresponding ions. The particle flux densities are determined by 
\begin{equation}
\label{j_a}
\bold{j}_a(\bold{r}_a,\bold{r}_b)=-\frac{1}{\gamma_a}c_{ab}(\bold{r}_a,\bold{r}_b)\nabla_a\bar{\mu}_{a}(\bold{r}_a,\bold{r}_b)+\frac{1}{\gamma_a}q_{a}c_{ab}(\bold{r}_a,\bold{r}_b)\left(\bold{E}-\nabla_a\bar{\psi}_b(\bold{r}_a,\bold{r}_b)\right),
\end{equation}
\begin{equation}
\label{j_b}
\bold{j}_b(\bold{r}_b,\bold{r}_a)=-\frac{1}{\gamma_b}c_{ba}(\bold{r}_b,\bold{r}_a)\nabla_b\bar{\mu}_{b}(\bold{r}_b,\bold{r}_a)+\frac{1}{\gamma_b}q_{b}c_{ba}(\bold{r}_b,\bold{r}_a)\left(\bold{E}-\nabla_b\bar{\psi}_a(\bold{r}_b,\bold{r}_a)\right),
\end{equation}
where we have introduced the two-point steady-state intrinsic chemical potentials, $\bar{\mu}_{a}(\bold{r}_a,\bold{r}_b)$, and 
\begin{equation}
\bar{\psi}_b(\bold{r}_a,\bold{r}_b)=\int \text{d}\bold{r}'\varrho(\bold{r}_{a}-\bold{r}')\psi_b(\bold{r}',\bold{r}_b)=\varrho \psi_b(\bold{r}_a,\bold{r}_b)
\end{equation}
is the smeared electrostatic potential at point $\bold{r}_a$ when an ion with total charge $q_b$ is located at point $\bold{r}_b$. The first term in equations (\ref{j_a}) and (\ref{j_b}) describes the diffusion of ions $a$ even in the absence of an external field. The second term describes the particle flux density due to the action of electrostatic forces from the external field and the 'molecular' electric field $-\nabla_a\bar{\psi}_b$ created at point $\mathbf{r}_a$ by the distorted ionic cloud, assuming that ion $b$ is located at point $\mathbf{r}_b$. The local electrostatic potential, $\psi_{b}=\psi_b(\bold{r}_a,\bold{r}_b)$, satisfies the standard Poisson equation
\begin{equation}
\label{Poisson}
\nabla^2\psi_b(\bold{r}_a,\bold{r}_b)=-\frac{1}{\varepsilon\varepsilon_0}\left[\sum\limits_{s}\rho_s(\bold{r}_a,\bold{r}_b)+q_b\varrho(\bold{r}_a-\bold{r}_b)\right],
\end{equation}
where $\varepsilon$ is the dielectric constant of medium, $\varepsilon_0$ is the vacuum permittivity, and we have introduced smeared ionic charge densities
\begin{equation}
\rho_s(\bold{r}_a,\bold{r}_b)=q_s\varrho\bar{c}_{sb}(\bold{r}_a,\bold{r}_b)=q_s\int \text{d}\bold{r}'\varrho(\bold{r}_a-\bold{r}')\bar{c}_{sb}(\bold{r}',\bold{r}_b).
\end{equation}
In what follows, we will use the following short-hand notation
\begin{equation}
\varrho f(\bold{r})=\int \text{d}\bold{r}'\varrho(\bold{r}-\bold{r}')f(\bold{r}').
\end{equation}

Now, we assume that the two-point non-equilibrium chemical potentials, $\bar{\mu}_a$, are dependent on the coordinates through the two-point concentrations, $c_{ab}(\mathbf{r}_a, \mathbf{r}_b)$, of the species, as for the corresponding equilibrium quantities:
\begin{equation}
\bar{\mu}_{a}(\bold{r}_a,\bold{r}_b)=\bar{\mu}_{a}(\{c_{lb}(\bold{r}_a,\bold{r}_b)\}).
\end{equation}
The latter assumption is known as the local equilibrium condition~\cite{landau2013fluid,kondepudi2014modern}. Thus, assuming weak deviations from inhomogeneity of solution, i.e. that $c_{lb}(\bold{r}_a,\bold{r}_b)=c_l+\bar{c}_{lb}(\bold{r}_a,\bold{r}_b)$, with concentration perturbations $\bar{c}_{lb}\ll c_l$,
we obtain in the linear approximation in $\bar{c}_{ab}$
\begin{equation}
\bar{\mu}_{a}(\bold{r}_a,\bold{r}_b)=\mu_a+\sum\limits_{l}\frac{\partial \mu_{a}}{\partial c_l}\bar{c}_{lb}(\bold{r}_a,\bold{r}_b)=\mu_a+\sum\limits_{l}J_{al}\bar{c}_{lb}(\bold{r}_a,\bold{r}_b),
\end{equation}
where $J_{al}=\partial \mu_a /\partial c_l$, with $\mu_a=\mu_a(\{c_{l}\})$ being the equilibrium bulk chemical potentials of species which will be specified below. Note also that due to translation invariance all functions should depend only on the vector $\bold{r}=\bold{r}_a-\bold{r}_b$, so that
\begin{equation}
\nabla_a=\nabla_r=\nabla,~\nabla_b=-\nabla,~\bar{c}_{ab}(\bold{r})=\bar{c}_{ba}(-\bold{r}).
\end{equation}
Considering all these assumptions in equations (\ref{j_a}), (\ref{j_b}) and (\ref{continous}) we arrive at the following basic electrokinetic equation in the linear approximation for $\bar{c}_{ab}$ and $\psi_b$
\begin{multline}
\label{kinetic_}
\sum\limits_l\left(\frac{c_a}{\gamma_a}J_{al}\nabla^2\bar{c}_{lb}(\bold{r})+\frac{c_b}{\gamma_b}J_{bl}\nabla^2\bar{c}_{la}(-\bold{r})\right)+\frac{q_a c_a}{\gamma_a}\nabla^2\bar{\psi}_b(\bold{r})+\frac{q_b c_b}{\gamma_b}\nabla^2\bar{\psi}_a(-\bold{r})\\=\bold{E}\cdot\left(\frac{q_a}{\gamma_a} \nabla \bar{c}_{ab}(\bold{r})-\frac{q_b}{\gamma_b} \nabla \bar{c}_{ba}(-\bold{r})\right).
\end{multline}
We call eq. (\ref{kinetic_}) {\sl modified Onsager-Fuoss (OF) equation}. Eq. (\ref{Poisson}) can be rewritten as
\begin{equation}
\label{Poisson_2}
\nabla^2\psi_b(\bold{r})=-\frac{1}{\varepsilon\varepsilon_0}\left[\sum\limits_{s}q_s\varrho\bar{c}_{sb}(\bold{r})+q_b\varrho(\bold{r})\right],
\end{equation}
where we have used the electroneutrality condition for the bulk solution $\sum_{a}q_a c_a=0$. Eq. (\ref{Poisson_2}) can be considered as closure for modified OF equation.

The system of coupled equations (\ref{kinetic_}) and (\ref{Poisson_2}) form the basis of the modified DHO theory with account for the nonlocal ionic charge distribution, which we will use further to calculate the relaxation contribution to ion mobilities and electrical conductivity.

\textbf{Onsager-Fuoss approach.} In accordance with the Onsager-Fuoss approach\cite{fuoss1978review,pitaevskii2012physical}, {considering only the case of very small external electric field}, we find the solutions in the form
\begin{equation}
\psi_{b}(\bold{r})=\psi_{b}^{(0)}(\bold{r})+\psi_{b}^{(1)}(\bold{r}),
\end{equation}
\begin{equation}
\bar{c}_{ab}(\bold{r})=\bar{c}_{ab}^{(0)}(\bold{r})+\bar{c}_{ab}^{(1)}(\bold{r}),
\end{equation}
where $\psi_{b}^{(0)}(\bold{r})$ and $\bar{c}_{ab}^{(0)}(\bold{r})$ are the electrostatic potential and 'perturbation' concentrations which is determined by equilibrium state at $\bold{E}=0$, described by the chemical equilibrium conditions~\cite{budkov2022modified} which determine the equilibrium intrinsic chemical potentials of species, i.e.
\begin{equation}
\label{chem_eq}
\bar{\mu}_{a}(\{\bar{c}_{lb}^{(0)}\})+q_{a}\bar{\psi}_b^{(0)}=\mu_a.
\end{equation}
Assuming rather small concentration perturbations, i.e. that $\bar{c}_{ab}^{(0)}\ll c_a$, eq. (\ref{chem_eq}) yields
\begin{equation}
\sum\limits_{l}J_{al}\bar{c}_{lb}^{(0)}(\bold{r})=-q_a\varrho\psi_b^{(0)}(\bold{r})
\end{equation}
or
\begin{equation}
\label{eq_Poisson}
\bar{c}_{ab}^{(0)}(\bold{r})=-\sum\limits_{l}\hat{J}_{al}q_l\varrho\psi_b^{(0)}(\bold{r}),
\end{equation}
where we introduced the reciprocal matrix, $\hat{J}_{ab}=(J^{-1})_{ab}$. {Therefore, we will be working within the framework of the linear electrostatic approximation, as well as the original DHO theory.} The equilibrium electrostatic potential can be obtained from equation
\begin{equation}
\nabla^2\psi_b^{(0)}(\bold{r})=-\frac{1}{\varepsilon\varepsilon_0}\left[\sum\limits_{s}q_s\varrho\bar{c}_{sb}^{(0)}(\bold{r})+q_b\varrho(\bold{r})\right],
\end{equation}
which can be rewritten with the help of eq. (\ref{eq_Poisson}) in the form
\begin{equation}
\label{DH}
\nabla^2\psi_b^{(0)}(\bold{r}) -\mathcal{K} \psi_b^{(0)}(\bold{r}) =-\frac{1}{\varepsilon\varepsilon_0}q_b\varrho(\bold{r}),
\end{equation}
where $\mathcal{K}=\kappa^2\varrho^2$ is the integral operator and
\begin{equation}
\label{kappa}
\kappa^2=\frac{1}{\varepsilon\varepsilon_0}\sum\limits_{ab}q_{a}\hat{J}_{ab}q_{b}=\frac{1}{\varepsilon\varepsilon_0}(q\hat{J}q).
\end{equation}
The solution of eq. (\ref{DH}) can be expressed in the form
\begin{equation}
\psi_b^{(0)}(\bold{r})=q_b G(\bold{r}),
\end{equation}
where function $G(\bold{r})$ is the solution of equation
\begin{equation}
\label{DH_2}
\nabla^2 G(\bold{r}) -\mathcal{K}  G(\bold{r}) =-\frac{1}{\varepsilon\varepsilon_0}\varrho(\bold{r}).
\end{equation}
 The solution of eq. (\ref{DH_2}) can be found by Fourier method with the following substitutions $\nabla\rightarrow i\bold{k}$, $\nabla^2\rightarrow -\bold{k}^2$, $\varrho f(\bold{r})\rightarrow \varrho_{-\bold{k}} f_{\bold{k}}$, $\varrho^2 f(\bold{r})\rightarrow |\varrho_{\bold{k}}|^2 f_{\bold{k}}$ that yields for Fourier component of $G(\bold{r})$
\begin{equation}
G_{\bold{k}}=\frac{1}{\varepsilon\varepsilon_0}\frac{\varrho_{\bold{k}}}{\bold{k}^2+\kappa^2|\varrho_{\bold{k}}|^2},
\end{equation}
where $\varrho_{\bold{k}}$ is the Fourier-image of function $\varrho(\bold{r})$ (charge form factor).
 
For point-like charges {for which $\varrho(\bold{r})=\delta(\bold{r})$ (or $\varrho_{\bold{k}}=1$)}, we obtain
\begin{equation}
G(\bold{r})=\frac{1}{4\pi \varepsilon\varepsilon_0}\frac{e^{-\kappa r}}{r},
\end{equation}
where  $r=|\bold{r}|$. Thus, $\kappa^{-1}$ in this case can be interpreted as the screening length~\cite{maggs2016general}. Its dependence on concentrations (and temperature) is determined by the dependence of the bulk chemical potentials on concentrations (and temperature), i.e. $\mu_{a}=\mu_{a}(\{c_{l}\})$ which we will specify below. The equilibrium perturbation concentrations, thereby, are
\begin{equation}
\label{c0}
\bar{c}_{ab}^{(0)}(\bold{r})=-q_b (\hat{J}q)_a \varrho G(\bold{r}),    
\end{equation}
where $(\hat{J}q)_a=\sum_l \hat{J}_{al}q_l$.

It is worth noting that in the case of point-like charges eqs. (\ref{DH}) and (\ref{kappa}) indicate that the present modified DHO theory is consistent with equilibrium theory based on the modified Poisson-Boltzmann equation~\cite{maggs2016general,budkov2022modified}.

The steady-state perturbations $\bar{c}_{ab}^{(1)}(\bold{r})$ and $\psi_b^{(1)}(\bold{r})$ which are the scalar values have the structure $(\bold{E}\cdot \bold{r}) f(r)$, where $f(r)$ is the function only of the absolute value of $\bold{r}$. Therefore, $\bar{c}_{ab}^{(1)}(\bold{r})$ and $\psi_b^{(1)}(\bold{r})$ are odd functions. Bearing in mind condition (\ref{trans}), we arrive at the following property
\begin{equation}
\bar{c}_{ab}^{(1)}(\bold{r})=\bar{c}_{ba}^{(1)}(-\bold{r}),
\end{equation}
which yields the following antisymmetry condition
\begin{equation}
\bar{c}_{ab}^{(1)}(\bold{r})=-\bar{c}_{ba}^{(1)}(\bold{r}).
\end{equation}
The latter results in $\bar{c}_{aa}^{(1)}(\mathbf{r}) = 0$. This, in turn, means that the concentration perturbations due to the application of an external field only occurs for pairs of different ions. Thus, we can derive the following system of equations for the steady-state variables, $\bar{c}_{ab}^{(1)}(\bold{r})$ and $\psi_a^{(1)}(\bold{r})$:
\begin{multline}
\label{kinetic}
\sum\limits_l\left(\frac{c_a}{\gamma_a}J_{al}\nabla^2\bar{c}_{lb}^{(1)}(\bold{r})+\frac{c_b}{\gamma_b}J_{bl}\nabla^2\bar{c}_{al}^{(1)}(\bold{r})\right)+\frac{q_a c_a}{\gamma_a}\nabla^2\bar{\psi}_b^{(1)}(\bold{r})-\frac{q_b c_b}{\gamma_b}\nabla^2\bar{\psi}_a^{(1)}(\bold{r})\\=-q_aq_b\left(\frac{(\hat{J}q)_a}{\gamma_a} -\frac{(\hat{J}q)_b}{\gamma_b}\right)\bold{E}\cdot\nabla (\varrho G(\bold{r})),
\end{multline}
\begin{equation}
\label{Poisson_Steady}
\nabla^2\psi_b^{(1)}(\bold{r}) =-\frac{1}{\varepsilon\varepsilon_0}\sum\limits_{l}q_l \varrho\bar{c}_{lb}^{(1)}(\bold{r}).
\end{equation}

\textbf{Application to one-component electrolyte solutions.} As an application of the proposed theory, let us consider only the case of one-component electrolyte solutions for which indices $a,b=1,2$, so that $q_1=z_1q$, $q_2=-z_2q$, $z_a=|q_a|/q$ is the charge number of ion $a$, $q=e$ is the elementary charge. In this case, only one variable $\bar{c}_{12}^{(1)}(\bold{r})\neq 0$. 
Thus, solving eqs. (\ref{kinetic}) and (\ref{Poisson_Steady}) by Fourier-method (for derivation details, see Appendix A), we arrive at the following general expression for relaxation contribution to conductivity
\begin{multline}
\label{sigma_r}
\Delta \sigma_r=-\frac{(z_1z_2)^3q^4\kappa^2}{18\pi\eta \varepsilon\varepsilon_0R}\frac{z_2 R_1 J_{11}+z_1 R_2 J_{22}+(z_1 R_1+z_2R_2)J_{12}}{(z_1^2J_{22}+z_2^2J_{11}+2z_1z_2J_{12})(z_2R_2 J_{11}+z_1R_1 J_{22})}\\\times\int\frac{\text{d}\bold{k}}{(2\pi)^3}\frac{|\varrho_{\bold{k}}|^2}{(\bold{k}^2+\kappa^2|\varrho_{\bold{k}}|^2)(\bold{k}^2+\kappa_1^2|\varrho_{\bold{k}}|^2)},
\end{multline}
where, as it follows from eq. (\ref{kappa}),
\begin{equation}
\label{kappa^2}
\kappa^2=\frac{q^2}{\varepsilon\varepsilon_0 |J|}(z_1^2J_{22}+z_2^2J_{11}+2z_1z_2J_{12}),
\end{equation}
and we took into account that $\hat{J}_{11}=J_{22}/|J|$, $\hat{J}_{22}=J_{11}/|J|$, $\hat{J}_{12}=-J_{12}/|J|$, $|J|=J_{11}J_{22}-J_{12}^2$. We introduced the following short-hand notation
\begin{equation}
\kappa_1^2=\frac{q^2z_1z_2}{\varepsilon\varepsilon_0}\frac{z_1R_2+z_2R_1}{z_2J_{11}R_2+z_1J_{22}R_1}
\end{equation}
and the 'reduced' Stokes radius $R=R_1R_2/(R_1+R_2)$, and also used the electroneutrality condition $z_1c_1=z_2c_2$ to exclude bulk concentrations of ions in the final expression. Eq. (\ref{sigma_r}) is one of the main results of this paper. {To apply it to the calculation of the relaxation contribution to conductivity, we need to specify the charge form factor $\varrho_{\mathbf{k}}$ and the matrix elements, $J_{ab} = \partial \mu_a / \partial c_b$. These quantities will be defined below.}

{It is instructive to consider the expression for the relaxation contribution, which follows from equation (\ref{sigma_r}), for the case when ions are modeled as point charges, i.e., when $\varrho_{\bold{k}} = 1$. By taking the integral, we can arrive at
\begin{equation}
\label{sigma_r_point}
\Delta \sigma_r=-\frac{q^4\kappa\,(z_1z_2)^3}{72\pi^2\eta \varepsilon\varepsilon_0R\left(1+\frac{\kappa_1}{\kappa}\right)}\frac{z_2R_1 J_{11}+z_1R_2 J_{22}+J_{12}(z_2R_2+z_1R_1)}{(z_1^2J_{22}+z_2^2J_{11}+2z_1z_2J_{12})(z_1R_1 J_{22}+z_2R_2 J_{11})}.
\end{equation}}

\section{Electrophoretic contribution}
Now, we turn to calculating the electrophoretic contribution for the total conductivity of an electrolyte solution. Let us consider a certain ion $b$ within the solution and the ionic atmosphere, surrounding it. Within this ionic cloud the electroneutrality condition is violated, so that its local charge density is
\begin{equation}
\bar{\rho}_{b}(\bold{r})=\sum\limits_{a}q_a \bar{c}_{ab}(\bold{r}).
\end{equation}
In the external field, $\bold{E}$, the solvent, which carries this cloud, experiences the volume force with density $\bold{f}_b(\bold{r})=\bar{\rho}_{b}(\bold{r})\bold{E}$.

Due to rather slow motion under external field, the solvent can be considered as incompressible that can be expressed via a well-known condition
\begin{equation}
\label{continous_eq}
\nabla\cdot \bold{v}_b(\bold{r})=0,
\end{equation}
where $\bold{v}_b$ is the velocity field around the ion $b$. The same reason allows us to ignore the inertial term in the Navier-Stokes equation and write it as follows~\cite{pitaevskii2012physical,avni2022conductance}
\begin{equation}
\label{stokes}
\eta\nabla^2 \bold{v}_b -\nabla P_{b}+\bold{f}_b=0,
\end{equation}
where $P_b$ is the local pressure in solvent around ion $b$. Using the Fourier-method (for details, see Appendix B), we can obtain the following expression for the electrophoretic contribution to the electrical conductivity
\begin{equation}
\label{sigma_ep}
\Delta\sigma_e=-\frac{4Iq^2\kappa^2}{3\eta}\int\frac{\text{d}\bold{k}}{(2\pi)^3}\frac{|\varrho_{\bold{k}}|^2}{\bold{k}^2(\bold{k}^2+\kappa^2|\varrho_{\bold{k}}|^2)},
\end{equation}
where $I=\sum_a z_a^2 c_a /2$ is the ionic strength. Eq. (\ref{sigma_ep}) is the second main result of this paper.

{As with the relaxation contribution to total conductivity (see equation (\ref{sigma_r})), in order to apply this to real systems, we need to specify the charge form factor $\varrho_\mathbf{k}$ and matrix elements $J_{ab}$ (in order to calculate the screening constant $\kappa$). 
This quantities will be provided below. Note that for the point-like charge model eq. (\ref{sigma_ep}) yields
\begin{equation}
\label{sigma_ep_point}
\Delta\sigma_{e}=-\frac{q^2I\kappa}{3\pi\eta}.
\end{equation}}

\section{Kohlrausch's law and beyond}
Based on the above results, we can write the total conductivity as
\begin{equation}
\label{sigma_tot}
\sigma =\sigma_0 +\Delta\sigma_{r}+\Delta\sigma_{e},
\end{equation}
where $\sigma_0={q^2}\left(z_1^2c_1/R_1+z_2^2c_2/R_2\right)/(6\pi\eta)$ is the Nernst-Einstien conductivity of very dilute solution, and the second and third terms are determined by eqs. (\ref{sigma_r}) and (\ref{sigma_ep}), respectively.

In the case of a very dilute ($c_a\to 0$) one-component 1:1 electrolyte solution, where we can approximate the bulk chemical potentials by the ideal-gas expressions
\begin{equation}
\mu_{a}=k_{B}T\ln(c_a \lambda_a^3/\xi_a),
\end{equation}
where $k_{B}$ is the Boltzmann constant, $T$ is the temperature, $\lambda_a$ is the thermal wavelength of the ion $a$, $\xi_a$ is the internal partition function of an ion $a$ (which drops out from the final expressions),  we obtain $J_{11}=J_{22}=k_{B}T/c$, $J_{12}=J_{21}=0$, where $c=c_1=c_2$. Thus in the case of point-like charge when $\varrho_{\bold{k}}=1$ after calculations of integrals, as expected, the developed theory results in the Debye-H\"{u}ckel-Onsager expression~\cite{onsager1927theory,onsager2002irreversible,fuoss1978review} for total conductivity of 1:1 electrolyte solution (so-called Kohlrausch's law~\cite{debye1987,fuoss1978review})
\begin{equation}
\sigma =\sigma_0(1-\xi l_{B}^{3/2}c^{1/2}),
\end{equation}
where in this case $\sigma_0={q^2c}/(6\pi\eta R)$ and $\xi$ is the numerical coefficient which has the following form:
\begin{equation}
\xi =\frac{2\sqrt{\pi}}{3}(\sqrt{2}-1)+\frac{4\sqrt{2\pi}R}{l_{B}},
\end{equation}
where $l_{B}=q^2/(4\pi\varepsilon\varepsilon_0 k_{B}T)$ is the standard Bjerrum length.

It is clear that Kohlrausch's law, which was originally obtained for very dilute solutions~\cite{onsager1927theory}, does not take into account the excluded volume effects of ions and their short-range specific interactions. To take into account the short-range specific interactions of ions, which should be significant in rather concentrated solutions, we can use the following expression for the free energy density~\cite{mazurunderstanding}:
\begin{equation}
\label{f}
f\left(\left\{c_{a}\right\}\right)=f_{id}\left(\left\{c_{a}\right\}\right)+f_{h s}\left(\left\{c_{a}\right\}\right)-\frac{1}{2} \sum_{ab} A_{ab} c_a c_b,
\end{equation}
where the first term describes the ideal gas contribution to the total free energy density
\begin{equation}
f_{id}=k_{B}T\sum\limits_{a}c_a\left(\ln(c_{a}\lambda_{a}^3/\xi_a)-1\right),
\end{equation}
the second term describes the contribution of steric interactions of the ions within the model of hard sphere mixture within the Percus-Yevick approximation~\cite{roth2010fundamental}:
\begin{equation}
\label{PY}
\frac{f_{hs}}{k_B T} = -n_0 \ln(1-n_3) + \frac{n_1 n_2}{1 - n_3} + \frac{n_2^3}{24\pi(1-n_3)^2}
\end{equation}
where $d_{a}$ is the diameter of ion $a$ (which should be interpreted as solvation diameter). The auxiliary variables are: $n_0 = \sum_{a}c_{a}$, $n_1 =  \sum_{a}c_ad_a/2$, $n_2 = \pi\sum_{a}c_ad_{a}^2$, $n_3 = \pi\sum_{a}c_{a}d_{a}^3/6$, $a=1,2$. The third term describes the contribution of the short-range specific interactions of ionic species with parameters, $A_{ab}$, of short-range specific interactions. The chemical potentials of species are $\mu_a=\partial f/\partial c_a$. Eq. (\ref{f}) can be used to calculate the matrix elements, $J_{ab}=\partial^2 f/\partial{c_a}\partial{c_{b}}$. We do not provide analytical formulas for these as they are too unwieldy. In practical calculations, it is more preferable to calculate the matrix elements $J_{ab}$ numerically rather than using their cumbersome analytical expressions.

Note that in this paper, we will only consider the case of aqueous electrolyte solutions, for which we assume that the parameters of short-range specific ion interactions are zero ($A_{ab}=0$). Therefore, we only take into account the effect of excluded volume (steric) interactions. However, a short-range specific interactions are important for room-temperature ionic liquids~\cite{goodwin2017mean,budkov2018theory}.

\section{Model charge form factor}
To calculate the relaxation and electrophoretic contributions to conductivity, we need to specify the charge form factor. We will use the following exponential {(Slater-type)} distribution function: 
\begin{equation}
\varrho(\bold{r})=\frac{1}{8\pi a^3}\exp\left[-{r}/{a}\right]
\end{equation}
with the following Fourier-image
\begin{equation} 
\label{varrho}
\varrho_{\bold{k}} = \int \text{d}\bold{r}e^{-i\bold{k}\cdot\bold{r}}\varrho(\bold{r})= \frac{1}{(1 + \bold{k}^{2}a^{2})^2},
\end{equation} 
where $a$ is the characteristic length scale of charge smearing, which should be of the order of the solvated ion size. As we demonstrate below, the use of this phenomenological distribution function results in a good approximation of the electrical conductivity of different electrolyte solutions up to relatively high concentrations. {Note that the Slater-type charge distribution function has been used in recent papers to model charge overlapping in molecular systems within quantum mechanics/molecular mechanics (QM/MM)~\cite{ohrn2016method}, or electrostatic interactions in macromolecular systems using dissipative particle dynamics (DPD)~\cite{posel2014dissipative,mao2015modeling,hendrikse2024dpd}. This distribution function is one of the simplest and, at the same time, qualitatively accurately reflects the charge distribution in real multi-electron systems~\cite{ohrn2016method}.}

Substituting (\ref{varrho}) into eq. (\ref{sigma_r}) and introducing new integration variable $x=k/\kappa$, we obtain
\begin{equation}
\label{sigma_r_}
\Delta \sigma_r=-\frac{q^4\kappa\,(z_1z_2)^3}{36\pi^3\eta \varepsilon\varepsilon_0R}\frac{z_2R_1 J_{11}+z_1R_2 J_{22}+J_{12}(z_2R_2+z_1R_1)}{(z_1^2J_{22}+z_2^2J_{11}+2z_1z_2J_{12})(z_1R_1 J_{22}+z_2R_2 J_{11})}\Lambda(s,u),
\end{equation}
where $s=\kappa a$, $u=\kappa_1/\kappa$ and 
\begin{equation}
\label{Lambda}
\Lambda(s,u)=\int\limits_{0}^{\infty}\text{d}x\frac{x^2(1+s^2x^2)^4}{\left(x^2(1+s^2x^2)^4+1\right)\left(x^2(1+s^2x^2)^4+u^2\right)}.
\end{equation}
The latter function has the following asymptotic  behavior:
\begin{equation}
\label{lambda_asymptotic}
\Lambda(s,u)\simeq
\begin{cases}
\frac{\pi}{2(1+u)}, &s\ll 1,\\
\frac{\pi(\sqrt{5}+1)}{10\left(1+u\right)\left(1+u^{1/5}+u^{2/5}+u^{3/5}+u^{4/5}\right)}\frac{1}{s^{4/5}}, &s\gg 1.
\end{cases}
\end{equation}
The regime where $s\ll 1$ corresponds to the region of validity of the standard DHO theory, while the new regime where $s\gg 1$ is approximately realized for electrolyte solutions with trivalent cations ($z_1=3$), room temperature ionic liquids and molten salts (see discussion below).

With this charge form factor, we obtain the electrophoretic contribution to conductivity
\begin{equation}
\label{sigma_e__}
\Delta\sigma_{e}=-\frac{2Iq^2\kappa}{3\pi^2\eta}\theta(s),
\end{equation}
where
\begin{equation}
\theta(s)=\int\limits_{0}^{\infty}\frac{\text{d}x}{1+x^2(1+x^2s^2)^4}.
\end{equation}

As well as in the case of function $\Lambda(s,u)$, we obtain the following asymptotic behavior for the auxiliary function $\theta(s)$ 
\begin{equation}
\label{theta_asymptotic}
\theta(s)\simeq
\begin{cases}
\frac{\pi}{2}, &s\ll 1,\\
\frac{\pi(\sqrt{5}+1)}{10}\frac{1}{s^{4/5}}, &s\gg 1.
\end{cases}
\end{equation}

\section{Numerical results}
Let us consider aqueous one-component electrolyte solutions with different ratios $z_1$:$z_2$ and compare their theoretically computed electrical conductivities with available experimental data using a model where ions are treated as hard spheres with smeared charges. Our model accounts for the hydration of ions through their hydration diameters, $d_a$, and dependence of the dielectric constant on ion concentrations,
$\varepsilon=\varepsilon_{w}-\alpha_1c_1 -\alpha_2c_2$, where $\varepsilon_w$ is the dielectric constant of pure water and $\alpha_{1,2}$ are the dielectric decrements~\cite{nakayama2015differential,ben2011dielectric,zhao2013influence,mazurunderstanding} of ions. Additionally, we take into account the dependence of the electrolyte solution's viscosity on the salt concentration, approximating a variety of available experimental data \cite{holze2016electrochemistry,laliberte2004model,laliberte2007model,isono1984density,out1980viscosity,haynes2016crc} with a Barthel-like exponential relationship~\cite{barthel1998electrolyte,boroujeni2023estimation}.

\begin{figure}[H]
  \centering
  \begin{subfigure}[b]{0.48\linewidth}
    \centering\includegraphics[height=5.5 cm]{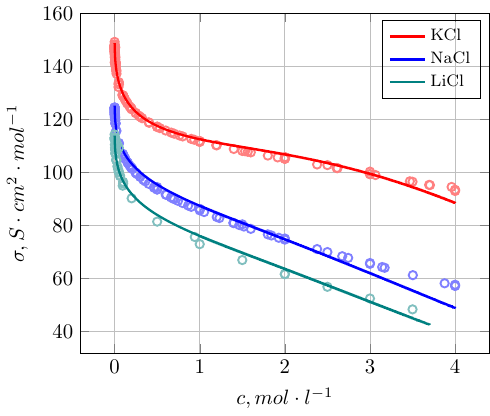}
    \caption{}\label{monovalent_chlorides}
  \end{subfigure}
  \hfill
  \begin{subfigure}[b]{0.48\linewidth}
    \centering\includegraphics[height=5.5 cm]{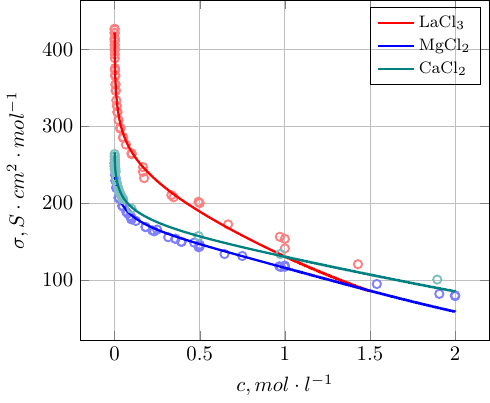}
    \caption{}\label{multivalent_chlorides}
  \end{subfigure}
    \begin{subfigure}[b]{0.48\linewidth}
    \centering\includegraphics[height=5.5 cm]{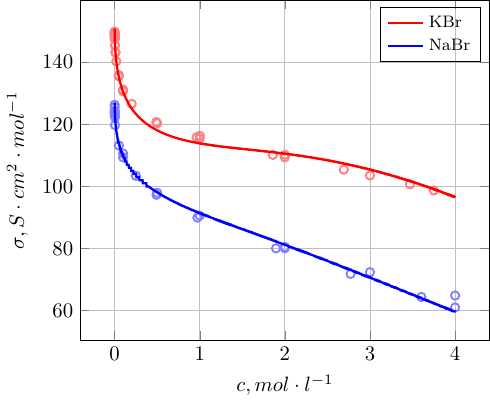}
    \caption{}\label{bromides}
  \end{subfigure}
  \hfill
    \begin{subfigure}[b]{0.48\linewidth}
    \centering\includegraphics[height=5.5 cm]{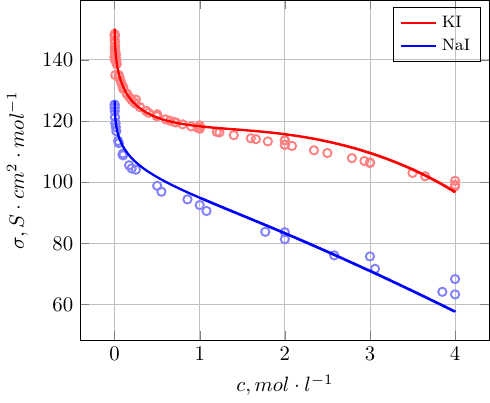}
    \caption{}\label{iodides}
  \end{subfigure}
  \begin{subfigure}[b]{0.48\linewidth}
    \centering\includegraphics[height=5.5 cm]{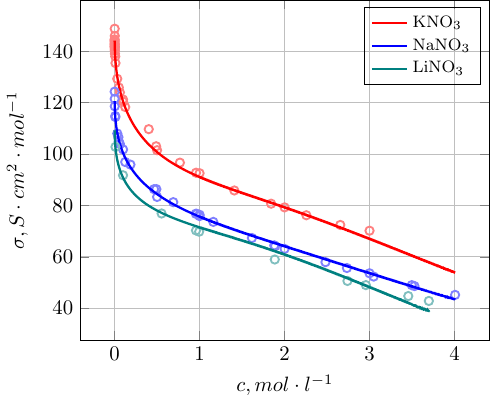}
    \caption{}\label{nitrates}
  \end{subfigure}
  \hfill
    \begin{subfigure}[b]{0.48\linewidth}
    \centering\includegraphics[height=5.5 cm]{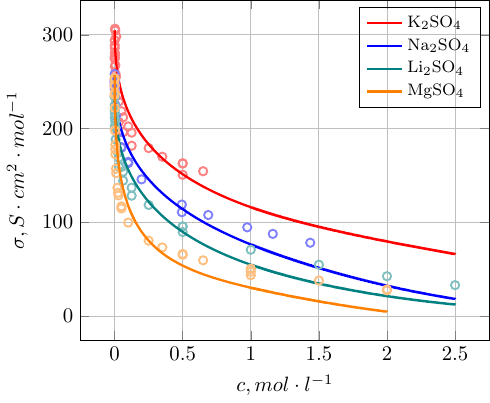}
    \caption{}\label{sulfates}
  \end{subfigure}
\caption{Molar electrical conductivity as a function of concentration for a set of aqueous electrolyte solutions at $T=298.15$ K. The lines represent the best fit of the experimental data \cite{holze2016electrochemistry,isono1984density,haynes2016crc,postler1970conductance,postler1970conductance2} using equations (\ref{sigma_tot}), (\ref{sigma_r_}) and (\ref{sigma_e__}).}
\label{fit_at_25}
\end{figure}

\begin{figure}[H]
\centering
\includegraphics[height=8 cm]{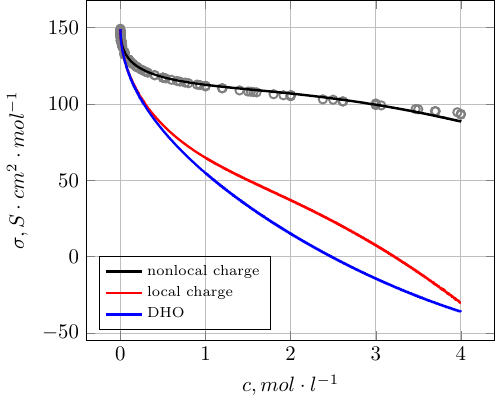}
\caption{{Comparison of the approximation obtained using the current theory, which takes into account the nonlocal nature of the ion charge, with the theory that does not take this into account ($a=0$), and with the standard Debye-H\"uckel-Onsager theory for aqueous KCl solution at $T=298.15$ K.}}
\label{contrib_comparison_3}
\end{figure}

Tables \ref{inpute_parameters} and \ref{fitting_parameters} present the values of the parameters used in the calculation of electrical conductivity and the fitting parameter -- the length scale of ionic charge smearing, $a$, -- respectively. Fig. \ref{fit_at_25} presents approximations of the experimental dependencies \cite{holze2016electrochemistry,isono1984density,haynes2016crc,postler1970conductance,postler1970conductance2,isono1984density,mccleskey2011electrical} of the molar electrical conductivity of a series of aqueous electrolyte solutions on concentration at $T=298.15$ K using the proposed model. As shown, the theory provides a decent approximation {(see the Table \ref{RAAD}  with RAAD $\%$ in Appendix C)} of the experimental data over a relatively wide concentration range, up to nearly 4 M for 1:1 electrolytes, with just one adjustable parameter. Despite the generally acceptable reproduction of the experimental data (see Fig. \ref{fit_at_25}), it is important to note that there are slight deviations from the reference data at high concentrations for most of studied electrolyte solutions, which will be discussed further below. {Note that Fig. \ref{contrib_comparison_3} shows a curve obtained using the "local" theory, i.e., for $a=0$ (see also eqs. (\ref{sigma_r_point}) and (\ref{sigma_ep_point})). As can be seen from Fig. \ref{contrib_comparison_3}, although accounting for the excluded volume and ignoring the non-local nature of charge does cause electrical conductivity to decrease with concentration more slowly than predicted by the Debye-H\"uckel-Onsager theory, this approach still gives incorrect results for concentrated solutions. Thus, simultaneous consideration of the excluded volume and the non-locality of the charge distribution is essential for the accurate description of the electrical conductivity of electrolytes over a wide range of concentrations.}

\begin{table}
    \centering
    \begin{tabular}{|l|c|c|c|}
        \hline
         ion & $d/2$  \cite{nightingale1959phenomenological}, \AA & $R_h$  \cite{nightingale1959phenomenological}, \AA & $\alpha$  \cite{hasted1948dielectric, harris1957dielectric}, $(mol/l)^{-1}$  \\
         \hline
         K$^+$              & 3.31 & 1.25 & 8  \\
         Na$^+$             & 3.58 & 1.84 & 8  \\
         Li$^+$             & 3.82 & 2.38 & 11 \\
         Ca$^{2+}$          & 4.12 & 3.10 & 27 \\
         Mg$^{2+}$          & 4.28 & 3.47 & 24 \\
         La$^{3+}$          & 4.52 & 3.96 & 35 \\
         Cl$^-$             & 3.32 & 1.21 & 3  \\
         Br$^-$             & 3.30 & 1.18 & 0  \\
         I$^-$              & 3.31 & 1.19 & 7  \\
         $\text{NO}_3^{-}$  & 3.35 & 1.29 & 0  \\
         $\text{SO}_4^{2-}$ & 3.79 & 2.30 & 7  \\
         \hline
    \end{tabular}
    \caption{List of input parameters of the model: hydrated radii, $d/2$, Stokes radii, $R_h$, and dielectric decrements, $\alpha$, of the ions at $T=298.15$ K.}
    \label{inpute_parameters}
\end{table}

\begin{table}
    \centering
    \begin{tabular}{|l|c|l|c|}
        \hline
         electrolyte & $a$, \AA & electrolyte & $a$, \AA \\
         \hline
         KCl        & 1.77  & NaI                       & 3.48  \\
         NaCl       & 2.53  & $\text{KNO}_3$            & 0.63  \\
         LiCl       & 2.32  & $\text{NaNO}_3$           & 1.39  \\
         CaCl$_2$   & 2.98  & $\text{LiNO}_3$           & 1.48  \\
         MgCl$_2$   & 3.04  & $\text{K}_2\text{SO}_4$   & 1.43  \\
         LaCl$_3$   & 3.59  & $\text{Na}_2\text{SO}_4$  & 1.68  \\
         KBr        & 1.38  & $\text{Li}_2\text{SO}_4$  & 1.46  \\
         NaBr       & 2.30  & $\text{Mg}\text{SO}_4$    & 2.11  \\
         KI         & 1.57  &                           &       \\
         \hline
    \end{tabular}
    \caption{Studied systems and corresponding characteristic length of charge smearing, $a$, obtained via the fit of the experimental data at $T=298.15$ K.}
    \label{fitting_parameters}
\end{table}

Let us now investigate whether the model can capture the effect of temperature changes on electrical conductivity. To achieve this, we need to account for the temperature dependence of the dielectric permittivity and viscosity of the electrolyte solution. We assume that temperature affects only the properties of the pure solvent, while concentration effects are considered independently, as described previously. The temperature dependence of the dielectric permittivity of water is described using a polynomial function \cite{malmberg1956dielectric}, and the viscosity dependence is determined via a simplified relation proposed in ref. \cite{laliberte2007model}.

\begin{figure}[H]
  \centering
  \begin{subfigure}[b]{0.48\linewidth}
    \centering\includegraphics[height=6 cm]{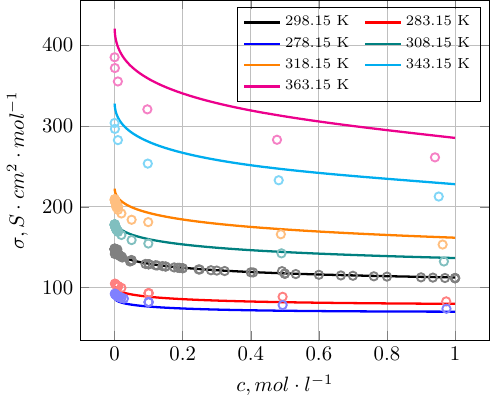}
    \caption{}\label{KCL_temp_dep_R_const}
  \end{subfigure}
  \hfill
  \begin{subfigure}[b]{0.48\linewidth}
    \centering\includegraphics[height=6 cm]{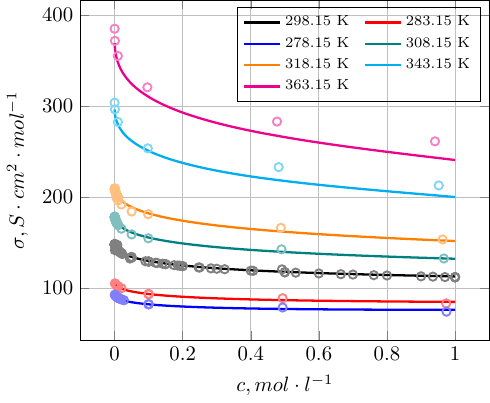}
    \caption{}\label{KCL_temp_dep_R}
  \end{subfigure}
\caption{Molar electrical conductivity isotherms of KCl aqueous solution. The lines represent the theoretical predictions with constant  (\subref{KCL_temp_dep_R_const}) and temperature-dependent (\subref{KCL_temp_dep_R}) Stokes radii. The symbols represent the experimental data~\cite{holze2016electrochemistry,mccleskey2011electrical,haynes2016crc}.}
\label{temperature_effect}
\end{figure}

If we consider the form factor as an essential characteristic of the ion, we should be enable to calculate electrical conductivity at various temperatures, provided we have obtained the value of parameter $a$ from preliminary experimental data fit. In Fig. \ref{temperature_effect}\subref{KCL_temp_dep_R_const}, we present the results of such an attempt for the KCl aqueous solution , i.e. we approximated the data at $T=298.15$ K and calculated conductivity with a fixed charge smearing parameter $a$ at different temperatures. While this method can estimate conductivity for temperatures close to the one used in fitting procedure, it fails at high temperatures, and, more importantly, gives inaccurate values of electrical conductivity at infinite dilution. On the other hand, the only parameters in the expression for $\sigma_0$, which can be considered as adjustable ones, are the Stokes radii of the ions, or the reduced radius $R$, introduced above, when dealing with a 1:1 electrolyte like KCl. Thus, an effective way to improve temperature-dependent conductivity calculations is to define these radii as functions of temperature. Although it is not straightforward to find experimental data for the conductivity at infinite dilution for each ion, in the case of KCl we can model both ions as identical due to their similar Stokes radii (see Table \ref{inpute_parameters}). By fitting the infinite dilution conductivity of KCl aqueous solution at different temperatures \cite{holze2016electrochemistry}, we obviously achieve improved agreement between experimental data and model calculations (see Fig. \ref{temperature_effect}\subref{KCL_temp_dep_R}).

\begin{figure}[H]
  \centering
  \begin{subfigure}[b]{0.48\linewidth}
    \centering\includegraphics[height=6 cm]{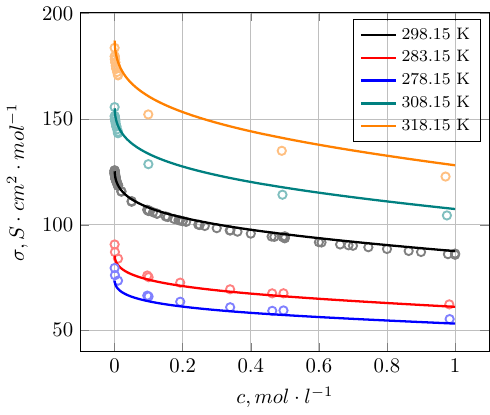}
    \caption{}\label{NaCL_temp_dep_R_const}
  \end{subfigure}
  \hfill
  \begin{subfigure}[b]{0.48\linewidth}
    \centering\includegraphics[height=6 cm]{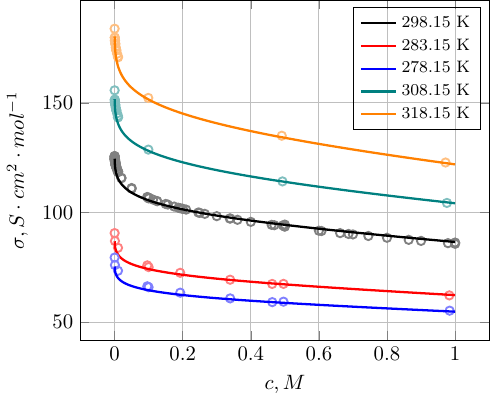}
    \caption{}\label{NaCL_temp_dep_R}
  \end{subfigure}
\caption{Molar electrical conductivity isotherms of NaCl aqueous solution. The lines represent the theoretical predictions with constant (\subref{NaCL_temp_dep_R_const}) and temperature-dependent (\subref{NaCL_temp_dep_R}) Stokes radii. The symbols represent the experimental data~\cite{holze2016electrochemistry,mccleskey2011electrical,haynes2016crc}.}
\label{temperature_effect_NaCl}
\end{figure}

Extending this approach to other aqueous electrolyte solutions appears straightforward: by considering the temperature dependence of the electrical conductivity at infinite dilution and assuming a constant Stokes radii ratio ($R_{Na}/R_{Cl}\approx1.5$), we achieve good agreement between theoretical and experimental isotherms for the case of NaCl aqueous solution (see Fig. \ref{temperature_effect_NaCl}\subref{NaCL_temp_dep_R_const} and \subref{NaCL_temp_dep_R}).

\section{Discussion}

As previously noted, the model results show slight deviations from experimental data at high concentrations for most of studied systems. The most significant discrepancies occur for multivalent ions, most notably, the anions -- the sulfates of lithium, sodium, potassium, magnesium. For these systems, the theory provides a reasonably good approximation only up to a concentration of approximately 0.3-0.5 mol/l. Minor contributions to observed deviations can be attributed to potential inaccuracies arising from experimental errors and the process of viscosity and electrical conductivity data fitting, as well as clear uncertainty related to the choice of dielectric decrement values~\cite{hasted1948dielectric,harris1957dielectric}. However, the primary cause of these deviations is likely due to the two main assumptions made in the utilized model. Firstly, the model neglects the formation of ionic pairs, which could significantly impact the results at rather high concentrations. Indeed, for monovalent electrolytes, ion pairs in a solution can form electrically neutral particles, which renormalize the total dielectric constant, whereas in the case of multivalent ones, ion pairs lead to the formation of ions with reduced charge. Thus, the contribution of excluded volume interactions cannot be accurately described within the hard spheres model used in this study. It may be more appropriate to represent ion pairs as dumbbells~\cite{boublik1990equation,gordievskaya2018interplay}. Using the expression for the total free energy of a mixture of convex bodies~\cite{boublik1974statistical} may provide a more accurate description of the steric interactions between these particles. Secondly, the model assumes identical form factors for both ions of the electrolyte, which may be an oversimplification, particularly for multivalent ions. Additionally, the model has limitation related to the concentration limit up to which conductivity can be accurately calculated. This limitation arises from the equation of state for hard spheres. Specifically, the ion concentrations must satisfy the condition $\pi\sum_{a}c_{a}d_{a}^3/6 < 1$  (see equation (\ref{PY})). For example, the maximum concentration at which this theory is applicable is approximately 4.2 mol/l for 1:1 electrolyte solutions. For 2:1, 1:2, and 3:1 electrolytes, the upper concentration limits are even lower. However, this limitation is not critical, as the developed model covers a sufficiently wide range of concentrations for all practically significant aqueous electrolyte solutions (see Fig. \ref{fit_at_25}).

\begin{figure}[H]
\centering
\includegraphics[height=8 cm]{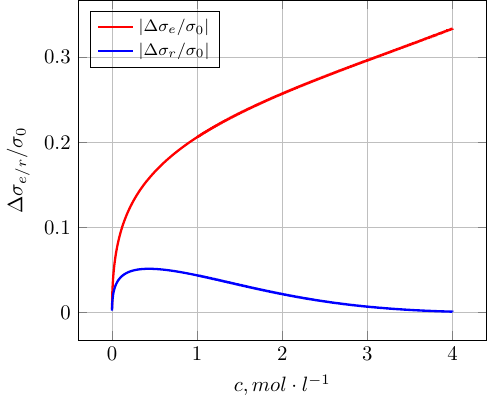}
\caption{The absolute values of the electrophoretic (red line) and relaxation (blue line) contributions to the electrical conductivity as functions of concentration. The values are reduced to the Nernst-Einstein conductivity, $\sigma_0$, for a KCl aqueous solution at $T=298.15$ K.}
\label{contrib_comparison}
\end{figure}

\begin{figure}[H]
  \centering
  \begin{subfigure}[b]{0.48\linewidth}
    \centering\includegraphics[height=6 cm]{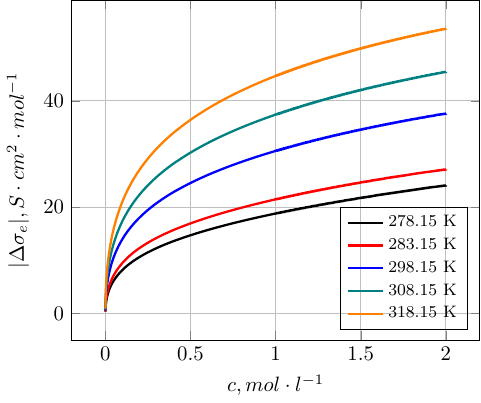}
    \caption{}\label{electrophortc_contrib_KCl_temp_dep}
  \end{subfigure}
  \hfill
  \begin{subfigure}[b]{0.48\linewidth}
    \centering\includegraphics[height=6 cm]{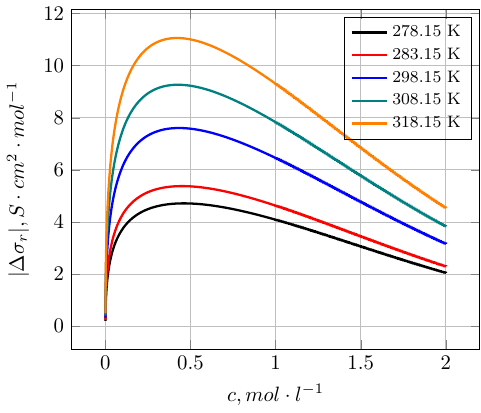}
    \caption{}\label{relaxation_contrib_KCl_temp_dep}
  \end{subfigure}
  \caption{The concentration dependencies of absolute values of the electrophoretic (\subref{electrophortc_contrib_KCl_temp_dep}) and relaxation (\subref{relaxation_contrib_KCl_temp_dep}) contributions to molar electrical conductivity for the case of KCl aqueous solution, plotted at different temperatures.}
\label{temperature_effect_2}
\end{figure}

Let us now compare the magnitudes of the electrophoretic and relaxation contributions to total conductivity and explore how they vary with changes in concentration and temperature. As shown in Fig. \ref{contrib_comparison}, the electrophoretic contribution is significantly larger than the relaxation one. This aligns with recent studies, which successfully described the conductivity of aqueous electrolyte solutions primarily by considering the electrophoretic effect~\cite{vinogradova2023conductivity,vinogradova2023electrophoresis}. The relaxation correction initially increases at low concentrations, peaks, and then declines towards zero. Note that the analogous behavior of the relaxation contribution to conductivity was predicted for the first time by Overbeck in reference \cite{jthg1943theorie}, and it was also predicted within the MSA framework in a recent paper~\cite{bernard2023analytical}. Small values of relaxation correction in concentrated solutions can be attributed to the strong binding of ions with their surroundings, which minimizes the distortion of the ionic atmosphere in the presence of an external electric field. In contrast, the electrophoretic correction steadily increases with concentration because a higher number of ions in the solution generates a stronger volume force on the solvent in an external field. Fig. \ref{temperature_effect_2} illustrates that both contributions to electrical conductivity increase with temperature. This is due to the reduction in electrostatic 'binding' between ions and their ionic atmospheres, which makes it easier for the ionic atmosphere to be distorted and carried away by an external field.

To obtain a precise estimate of the experimental data on electrical conductivity, it is important to consider the impact of electrolyte additives on the shear viscosity and dielectric constant of the solution. The literature provides this information for most electrolytes relevant to our research. Although neglecting these effects may affect accuracy, using the values of the pure solvent for these properties can still yield reasonably accurate estimates, as demonstrated in recent studies~\cite{vinogradova2023electrophoresis,vinogradova2023conductivity,avni2020charge}. It is important to note that the fits of the experimental data on electrical conductivity as a function of concentration were performed for systems with ions whose dielectric decrements are known. However, this is not a fundamental limitation, as in the cases when data on ion decrements are unavailable~\cite{boroujeni2023estimation}, one can use empirical formulas to describe the dependence of the dielectric constant of the aqueous electrolyte on concentration. In a similar manner, empirical formulas can also be applied to estimate the shear viscosity of aqueous electrolyte solutions.

{It is important to note that there seems to be linear correlation between the ion charge smearing parameter, $a$, and the mean square hydration radius, $\sqrt{(r_{+}^2+r_{-}^2)/2}$, of ions. As can be seen in Fig. \ref{correlation_rH's_and_a's} in Appendix C, this correlation is quite reliable for chlorides (coefficient of determination $\approx 0.91$).}

\begin{figure}[H]
\centering
\includegraphics[height=8 cm]{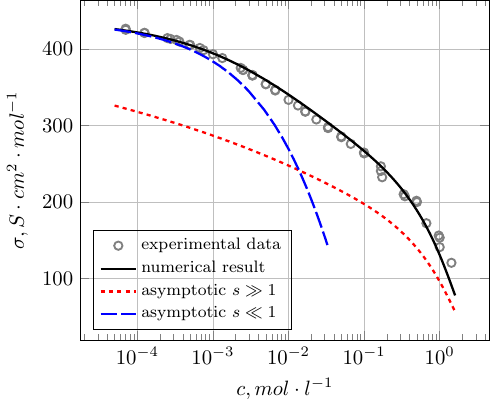}
\caption{Molar electrical conductivity of LaCl$_3$ aqueous solution at $T=298.15$ K (solid line) and its asymptotic behavior for $s=\kappa a \ll 1$ (dashed line) and $s=\kappa a \gg 1$ (dotted line), introduced in eqs. (\ref{lambda_asymptotic}) and (\ref{theta_asymptotic}).}
\label{asymptotics}
\end{figure}

It is interesting to note that asymptotic behavior of $\kappa a\gg 1$ (eqs.  (\ref{lambda_asymptotic}) and (\ref{theta_asymptotic})) can be approximately achieved for 3:1 aqueous electrolyte solutions at sufficiently high concentrations. As shown in Fig. \ref{asymptotics}, electrical conductivity for LaCl$_3$ aqueous solutions tends to this regime at concentrations of approximately 1-1.5 M. This regime could also be achieved for room-temperature ionic liquids and molten salts. However, for room-temperature ionic liquids, it is more relevant to use the previously mentioned equation of state for mixtures of convex bodies, such as dumbbells~\cite{boublik1990equation}, to account for the specific shape of molecular ions (e.g., the length of the alkyl chain of cations). Additionally, the formation of ionic clusters in room-temperature ionic liquids should be considered, as it can reduce the total electrical conductivity~\cite{feng2019free,avni2020charge,goodwin2022cracking}. Furthermore, on the basis of general considerations, approximating the unique charge form factor in the case of room-temperature ionic liquids would be quite rough, and defining these form factors would be a separate task. Nevertheless, it would be interesting to explore the potential of this theory for modeling the electrical conductivity of solutions containing room-temperature ionic liquids~\cite{bevster2016mobility,bevster2011association}. These topics could be the focus of future research.

\begin{figure}[H]
\centering
\includegraphics[height=8 cm]{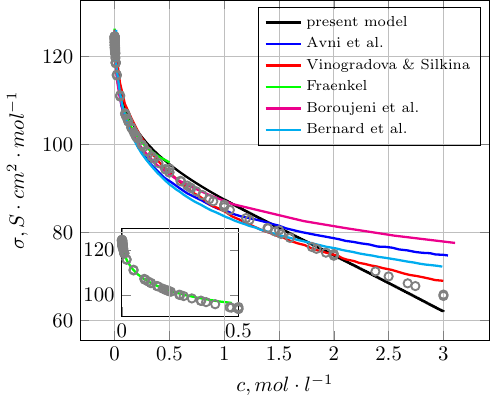}
\caption{{Comparison of the electrical conductivity predictions of the current theory with those of previously formulated models for NaCl aqueous solution at $T=298.15$ K. Data taken from: Avni et al. \cite{avni2022conductivity_}, Vinogradova-Silkina \cite{vinogradova2023electrophoresis}, Fraenkel \cite{fraenkel2018improved}, Boroujeni et al. \cite{naseri2023new}, Bernard et al. \cite{bernard2023analytical}. The dimensionless values of conductivity from refs. \cite{avni2022conductivity_,vinogradova2023electrophoresis} were multiplied by the experimental value \cite{holze2016electrochemistry} of the conductivity at infinite dilution of 126.5 $S\cdot cm^2\cdot mol^{-1}$. The inset emphasizes the comparison between Fraenkel's model predictions and the experiment.}}
\label{contrib_comparison_2}
\end{figure}

{It is beneficial to compare the predictions of the current theory with those mentioned in the Introduction by the example of aqueous NaCl solution at $T=298.15$ K. The models exhibit good agreement within the low to moderate concentration range (up to 0.5-1 mol/l). For highly concentrated solutions, the Vinogradova-Silkina model and the proposed model yield values that are more close to the experimental data (see Fig. \ref{contrib_comparison_2}).}

\section{Conclusion}
The mean field theory of electrolyte solutions, which is introduced in this paper as an extension of the classical Debye-H\"uckel-Onsager theory, takes into account the specific characteristics of ions, such as their steric interactions, hydration, and nonlocal charge distributions. This model provides analytical expressions for the mobilities of ions and the electrical conductivity, considering relaxation and hydrodynamic effects. At low concentrations, the theory agrees well with Kohlrausch's law. At higher concentrations, however, it shows good agreement with experimental data that deviate significantly from Kohlrausch's limiting law. This deviation can be attributed to various physical and chemical factors, including steric effects, ion hydration, and nonlocal charge distribution in solvated ions. By introducing a single adjustable parameter to describe the scale of the nonlocal charge distribution of solvated ions, and using experimental data on shear viscosity, Stokes radii, hydration radii, and dielectric decrements of ions, the model was able to successfully approximate experimental data on electrical conductivity of aqueous electrolyte solutions over a wide concentration range. These results provide predictive insights that are consistent with existing experimental findings. The theory can be easily expanded to estimate the electrical conductivity of binary electrolyte solutions.

\textbf{Data availability statement.} {\sl The data that support the findings of this study are available from the corresponding author upon reasonable request.}

\textbf{Acknowledgements.}
YAB would like to express its gratitude to O.I. Vinogradova for bringing to his attention some issues raised in the paper. YAB thanks A.A. Kornyshev for motivating discussions. The calculations were performed using computational resources from HPC facilities at NRU HSE. The article was prepared within the framework of the project 'Mirror Laboratories' HSE University. NNK and YAB would like to thank L.P. Safonova, A.M. Kolker, L.E. Shmukler, M.A. Krestyaninov, and D.M. Makarov for their assistance in finding the experimental data. NNK also appreciates S.E. Kruchinin's assistance with the numerical computations.

\appendix

\section{Derivation of eq. (\ref{sigma_r})}
In this Appendix, we provide the details of the derivation of the general expression for the relaxation contribution (\ref{sigma_r}) to the conductivity of one-component electrolyte solutions. The basic equations (\ref{kinetic}) and (\ref{Poisson_Steady}) for one-component electrolyte solutions take the form 
\begin{equation}
\label{1}
\nabla^2\psi_{2}^{(1)}(\bold{r})=-\frac{z_1q}{\varepsilon\varepsilon_0}\varrho\bar{c}_{12}^{(1)}(\bold{r}),
\end{equation}
\begin{multline}
\label{2}
\left(\frac{c_1 J_{11}}{\gamma_1}+\frac{c_2 J_{22}}{\gamma_2}\right)\nabla^2\bar{c}_{12}^{(1)}(\bold{r})-\frac{q^2}{\varepsilon\varepsilon_0}\left(\frac{z_1^2c_1}{\gamma_1}+\frac{z_2^2c_2}{\gamma_2}\right)\varrho^2\bar{c}_{12}^{(1)}(\bold{r})\\=z_1 z_2q^3\left(\frac{z_1\hat{J}_{11}-z_2\hat{J}_{12}}{\gamma_1}-\frac{z_1\hat{J}_{12}-z_2\hat{J}_{22}}{\gamma_2}\right)\bold{E}\cdot\nabla (\varrho G(\bold{r})).
\end{multline}
 The system of equations (\ref{1}) and (\ref{2}) can be easily solved by Fourier method. For Fourier-images $\bar{c}_{12}^{(1)}{_{\bold{k}}}$ and $\psi_{2}^{(1)}{_{\bold{k}}}$ the system of partial differential equations transforms into algebraic equations. The result for $\bar{c}_{12}^{(1)}{_{\bold{k}}}$ and $\psi_{2}^{(1)}{_{\bold{k}}}$ is
\begin{equation}
\bar{c}_{12}^{(1)}{_{\bold{k}}}=-\frac{z_1z_2q^3}{\varepsilon\varepsilon_0 |J|}\frac{\gamma_2(z_1J_{22}+z_2J_{12})+\gamma_1 (z_2J_{11}+z_1J_{12})}{c_1\gamma_2 J_{11}+c_2\gamma_1 J_{22}} \frac{i(\bold{k}\cdot \bold{E})\varrho_{\bold{k}}}{(\bold{k}^2+\kappa^2|\varrho_{\bold{k}}|^2)(\bold{k}^2+\kappa_1^2|\varrho_{\bold{k}}|^2)},
\end{equation}
\begin{equation}
\psi_{2}^{(1)}{_{\bold{k}}}=\frac{z_1q}{\varepsilon\varepsilon_0 \bold{k}^2}\varrho_{-\bold{k}}\bar{c}_{12}^{(1)}{_{\bold{k}}}.
\end{equation}

Since $\psi^{(1)}_{2}(\mathbf{r}_{1} - \mathbf{r}_2)$ is the additional electrostatic potential at point $\mathbf{r}_{1}$, which occurs when ion 2 is placed at point $\mathbf{r}_2$, then the corresponding electric field is 
\begin{equation} 
\mathbf{E}_{2}^{(1)}(\mathbf{r}_1-\mathbf{r}_2) = -\nabla_1 \psi^{(1)}_2(\mathbf{r}_1-\mathbf{r}_2) = -\nabla \psi_2^{(1)}(\mathbf{r}). 
\end{equation} 
At $\mathbf{r} = 0$, i.e., at $\mathbf{r}_1=\mathbf{r}_2$, this field gives the additional field acting on ion 2 and thereby changing its mobility. The Fourier-image $\mathbf{E}_{2}^{(1)}{_\bold{k}}=-i\bold{k}\psi_{2}^{(1)}{_{\bold{k}}}$. Therefore, 
\begin{equation}
\mathbf{E}_{2}^{(1)}(0)=-\int\frac{\text{d}\bold{k}}{(2\pi)^3}i\bold{k}\psi_{2}^{(1)}{_{\bold{k}}}.
\end{equation}
Using the substitution $\mathbf{k}(\mathbf{k}\cdot \mathbf{E}) \rightarrow \bold{E}k^2/3$ which occurs after averaging over the solid angle, we obtain
\begin{equation}
\mathbf{E}_{2}^{(1)}(0)=-\beta_2\bold{E},
\end{equation}
where 
\begin{equation}
\label{alpha2}
\beta_2=\frac{q^4z_1^2z_2}{3(\varepsilon\varepsilon_0)^2 |J|}\frac{\gamma_2(z_1J_{22}+z_2J_{12})+\gamma_1 (z_2J_{11}+z_1J_{12})}{c_1\gamma_2 J_{11}+c_2\gamma_1 J_{22}}\int\frac{\text{d}\bold{k}}{(2\pi)^3} \frac{|\varrho_{\bold{k}}|^2}{(\bold{k}^2+\kappa^2|\varrho_{\bold{k}}|^2)(\bold{k}^2+\kappa_1^2|\varrho_{\bold{k}}|^2)}.
\end{equation}
To obtain $\beta_1$, we have to swap indices 1 and 2 in eq. (\ref{alpha2}).

The total electric field acting on ion $2$ is
\begin{equation}
\bold{E}+\mathbf{E}_{2}^{(1)}(0)=(1-\beta_2)\bold{E}.
\end{equation}
Therefore, the mobility of the ion $a$ ($a=1,2$) is
\begin{equation}
m_a=\frac{1-\beta_a}{\gamma_a},
\end{equation}
so that relaxation contribution to ion mobility is
\begin{equation}
m_a^{(r)}=-\frac{\beta_a}{\gamma_a}.
\end{equation}
Thus, taking into account that $\Delta \sigma_r=\sum\limits_{a}q_{a}^2c_am_a^{(r)}$, after some algebra we arrive at eq. (\ref{sigma_r}).

\section{Derivation of eq. (\ref{sigma_ep})}
In this Appendix, we discuss the derivation of the general expression for the electrophoretic contribution (\ref{sigma_r}) to the conductivity of one-component electrolyte solutions. In Fourier-representation equations (\ref{continous_eq}) and (\ref{stokes}) can be rewritten as
\begin{equation}
\bold{k}\cdot\bold{v}_{b\bold{k}}=0,~-\eta k^2\bold{v}_{b\bold{k}}-i\bold{k}P_{b\bold{k}}+\bold{E}\bar{\rho}_b{_\bold{k}}=0,
\end{equation}
where $k=|\bold{k}|$. Multiplying scalarly the second equation by $i\bold{k}$ and using the first equation, after simple algebra we obtain
\begin{equation}
\bold{v}_{b\bold{k}}=\frac{\bar{\rho}_b{_\bold{k}}}{\eta} \frac{k^2\bold{E}-\bold{k}(\bold{k}\cdot \bold{E})}{k^4}.
\end{equation}
In the linear approximation in $\bold{E}$, we have
\begin{equation}
\label{vk}
\bold{v}_{b\bold{k}}\simeq \frac{\bar{\rho}_{b\bold{k}}^{(0)}}{\eta} \frac{k^2\bold{E}-\bold{k}(\bold{k}\cdot \bold{E})}{k^4}
\end{equation}
Taking into account eq. (\ref{c0}) in Fourier-representation, we obtain
\begin{equation}
\rho_{b\bold{k}}^{(0)}=-\frac{q_b\kappa^2|\varrho_{\bold{k}}|^2}{\bold{k}^2+\kappa^2|\varrho_{\bold{k}}|^2},
\end{equation}
Using eq. (\ref{vk}), we obtain the velocity of solvent at $\bold{r}=0$:
\begin{equation}
\bold{v}_b(0)=\int\frac{\text{d}\bold{k}}{(2\pi)^3}\bold{v}_{b\bold{k}}.
\end{equation}
This velocity should be added to velocity $q_b m_{b}^{(0)}\bold{E}$. Therefore, the electrophoretic contribution to the mobility 
\begin{equation}
m_a^{(e)}=-\frac{2\kappa^2}{3\eta}\int\frac{\text{d}\bold{k}}{(2\pi)^3}\frac{|\varrho_{\bold{k}}|^2}{\bold{k}^2(\bold{k}^2+\kappa^2|\varrho_{\bold{k}}|^2)}.
\end{equation}
Thus, taking into account that $\Delta\sigma_e = \sum_{a} q_a^2 c_a m_a^{(e)}$, we immediately obtain eq. (\ref{sigma_ep}).

\section{}

\begin{table}[H]
    \centering
    \begin{tabular}{|l|c|c|c|}
        \hline
         electrolyte & RAAD, \% & $c_{max},~mol\cdot l^{-1}$  & $N$ \\
         \hline
         KCl                        & 1.34   & 4      &   160  \\
         NaCl                       & 1.71   & 4      &   115  \\
         LiCl                       & 1.62   & 4      &   42   \\
         CaCl$_2$                   & 0.94   & 1.9    &   31   \\
         MgCl$_2$                   & 2.46   & 1.9    &   53   \\
         LaCl$_3$                   & 0.93   & 1.4    &   45   \\
         KBr                        & 0.62   & 3.8    &   31   \\
         NaBr                       & 1.34   & 4      &   24   \\
         KI                         & 0.88   & 4      &   74   \\
         NaI                        & 2.70   & 4      &   33   \\
        $\text{KNO}_3$              & 0.98   & 3      &   33   \\
        $\text{NaNO}_3$             & 1.20   & 4      &   32   \\
        $\text{LiNO}_3$             & 2.97   & 3.7    &   10   \\
        $\text{K}_2\text{SO}_4$     & 11.74  & 0.65   &   19   \\
        $\text{Na}_2\text{SO}_4$    & 15.51  & 1.4    &   25   \\
        $\text{Li}_2\text{SO}_4$    & 19.48  & 2.5    &   25   \\
        $\text{Mg}\text{SO}_4$      & 26.44  & 2      &   28   \\ 
        \hline
    \end{tabular}
    \caption{Relative Absolute Average Deviation (RAAD) of studied aqueous electrolyte solutions at 298.15 K; $c_{max}$ is the highest concentration within the studied range, $N$ is a number of experimental points, RAAD$ = \frac{1}{N}\sum\left(\frac{|\sigma_{exp}-\sigma_{calc}|}{\sigma_{exp}}\right)\times 100$.}
    \label{RAAD}
\end{table}

\begin{figure}[H]
\centering
\includegraphics[height=8 cm]{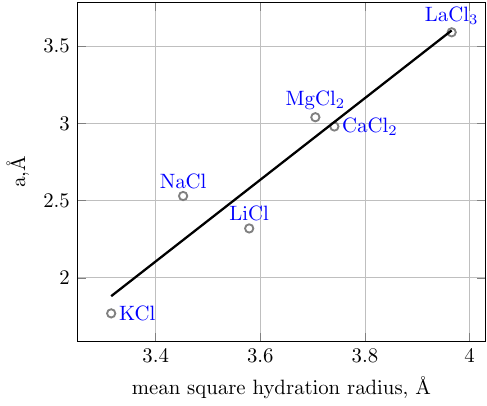}
\caption{Linear correlation between the ion charge smearing length, $a$, and the mean square hydration radius, $\sqrt{(r_{+}^2+r_{-}^2)/2}$, of the ions for the chlorides at 298.15 K. Coefficient of determination $\approx 0.91$.}
\label{correlation_rH's_and_a's}
\end{figure}

\selectlanguage{english}
\bibliography{name}

\end{document}